\documentclass[aps,prl,showpacs,nobibnotes,twocolumn,
nofootinbib,nobalancelastpage,superscriptaddress]{revtex4}
\usepackage{graphicx}
\usepackage{amsmath}
\usepackage{epstopdf}
\usepackage{latexsym}
\usepackage{dcolumn}
\usepackage{bm}
\input{epsf}
\newcommand{\be}{\begin{equation}}
\newcommand{\ee}{\end{equation}}
\newcommand{\ba}{\begin{eqnarray}}
\newcommand{\ea}{\end{eqnarray}}
\newcommand{\bi}{\begin{itemize}}
\newcommand{\ei}{\end{itemize}}
\newcommand{\ga}{\gtrsim}
\newcommand{\bfi}{\begin{figure}
\epsfxsize=9cm
\epsffile}
\newcommand{\efi}{\end{figure}}

\newcommand{\la}{\lesssim}

\begin{document}
\title{A discriminating probe of gravity at cosmological scales} 
\author{Pengjie Zhang}
\affiliation{Shanghai Astronomical Observatory, Chinese Academy of
  Science, 80 Nandan Road, Shanghai, China, 200030}
\affiliation{Joint Institute for Galaxy and Cosmology (JOINGC) of
SHAO and USTC}
\author{Michele Liguori}
\affiliation{Department of Applied Mathematics and Theoretical
  Physics, Centre for Mathematical Sciences, University of Cambridge,
  Wilberforce Road, Cambridge, CB3 0WA, United Kingdom}  
\author{Rachel Bean}
\affiliation{Department of Astronomy, Cornell University, Ithaca, NY 14853}
\author{Scott Dodelson}
\affiliation{Center for Particle Astrophysics,
Fermi National Accelerator Laboratory, Batavia, IL~~60510-0500}
\affiliation{Department of Astronomy \& Astrophysics, The University
  of Chicago, Chicago, IL~~60637-1433}
\email{pjzhang@shao.ac.cn} 
\begin{abstract}
The standard cosmological model is based on general relativity and
includes dark matter  and dark energy. An important prediction of this
model is a fixed relationship between the gravitational 
potentials responsible for gravitational lensing and the matter
overdensity. Alternative  theories of gravity often make different
predictions for this relationship. We propose a set of measurements
which can test the lensing/matter relationship, thereby  
distinguishing between dark energy/matter models and models in which
gravity differs from general relativity.  Planned optical, infrared
and radio galaxy and lensing surveys will be 
able to measure $E_G$, an observational quantity whose
expectation value is equal to the ratio of the Laplacian of the
Newtonian potentials to the  peculiar velocity divergence, to percent
accuracy.  We show that this  will easily  separate   alternatives
such as $\Lambda$CDM, DGP, TeVeS and $f(R)$ gravity. 
\end{abstract}
\maketitle
{ \bf Introduction}.---
Predictions based on general relativity plus the Standard Model of
particle physics  
are at odds with a variety of independent astronomical observations on
galactic and cosmological scales. This failure has led to
modifications in  particle physics.  By introducing dark matter and dark 
  energy, cosmologists have been able to account for a wide range of
  observations, from 
  the overall expansion of the universe to 
  the large scale structure of the early and late universe~\cite{Reviews}. 
Alternatively, attempts have been made to modify
general relativity at galactic~\cite{MOND} or cosmological
scales~\cite{DGP,fR}.   
A fundamental question then arises: {\it Can the two sets of 
modifications be distinguished from one another?}  

The answer is ``No''
if only the zero order expansion of the universe is considered. 
By allowing the dark energy equation of
state $w_{\rm DE}$ to be a free function, the expansion history $H(z)$
produced by any modified gravity can be mimicked exactly.  
Fortunately, structure formation in
modified gravities in general differs
\cite{Yukawa,Skordis06,Dodelson06,DGPLSS,Koyama06,fRLSS,Zhang06,Bean06,MMG,Uzan06,Caldwell07,Amendola07}
from that 
in general relativity. The difference we focus on here is the
relationship between gravitational potentials 
responsible for gravitational lensing and the 
matter overdensity. Lensing is sensitive to
$\nabla^2(\phi-\psi)$ along the line of sight  
where $\phi$ and $\psi$
are the two potentials in the perturbed Friedman-Robertson-Walker
metric: $ds^2=(1+2\psi)dt^2-a^2(1+2\phi)d{\bf 
  x}^2$ and $a$ is the scale factor. In standard general relativity (GR),
in the absence of anisotropic stresses, $\phi=-\psi$, 
  so lensing is sensitive to $\nabla^2\phi$. The Poisson equation 
  algebraically relates $\nabla^2\phi$ to the fractional
  overdensity $\delta$, so lensing is essentially determined by
  $\delta$ along the line of sight. 
  This is a prediction of the standard, GR-based theory that is
  generally not obeyed by alternate theories of gravity. 
  
  Testing this prediction is non-trivial. Astronomers often use the
  galaxy overdensity as a probe of the underlying 
  matter overdensity, but the two are not exactly equal. 
  Here we propose a test of this prediction which is relatively
  insensitive to the problem of galaxy bias. The basic idea 
  is simple: 
  \begin{itemize}
  \item Extract the matter overdensity at a given redshift by
  measuring the velocity field. Matter conservation 
  relates velocities to the overdensities. The measurement of the
  velocity field can be accomplished by studying the 
  anisotropy of the galaxy power spectrum in redshift space.
  \item Extract the lensing signal  at this redshift by
  cross-correlating these galaxies and lensing maps reconstructed from
  background galaxies.
  \end{itemize}

More quantitatively, the galaxy-velocity cross power spectrum
$P_{g\theta}\equiv -\langle \delta_g({\bf k})\theta(-{\bf k})\rangle$
can be inferred from redshift distortions in a galaxy distribution.
Here,
$\theta\equiv \nabla\cdot {\bf 
  v}/H(z)$ and  ${\bf v}$ is the comoving peculiar velocity.  In the
linear regime, matter conservation relates $\theta$ to  
$\delta$ by $\theta=-\dot{\delta}/H=-\beta \delta$, where
$\beta\equiv d\ln D/d\ln a$ and $D$ is the linear density growth factor.  
So, $P_{g\theta} = \beta P_{g\delta}$, satisfying the first goal
above. Cross correlating the same galaxies  with lensing maps
constructed from galaxies at higher redshifts,
$P_{\nabla^2(\phi-\psi)g}$ can be measured.  
The ratio of these two
cross-spectra therefore is a direct probe of
$\nabla^2(\phi-\psi)/(\beta\delta)$.   
It does not  
depend on galaxy bias or on the initial matter fluctuations,  at
least in the linear regime.  Modifications  
in gravity will in general leave signatures in either  $\beta$ and/or the
Poisson equation.

{\bf Galaxy-Velocity Cross-correlation}.---
A galaxy's peculiar motion shifts its apparent radial position from $x_z$ to
  $x_z^s=x_z+v_z/H(z)$ in redshift space, where $v_z$ is the comoving
radial peculiar velocity. The coherent velocity component
  changes 
  the galaxy number overdensity from $\delta_g$ to $\delta^s_g\simeq
  \delta_g-\nabla_z v_z/H(z)$. Galaxy random motions mix different
  scales and damps the power 
  spectrum on small scales. The redshift space galaxy power spectrum
  therefore has the general form  (\cite{Scoccimarro04} and references
  therein)  
\ba
\label{eqn:RD}
P^s_g({\bf
  k})=\left[P_g(k)+2u^2 P_{g\theta}(k)+u^4
P_{\theta}(k)\right]F\left(\frac{k^2u^2 \sigma^2_v}{H^2(z)}\right)
\ea
where $u=k_{\parallel}/k$ is the cosine of the angle of the ${\bf k}$
vector with 
respect to radial direction; $P_g$, $P_{g\theta}$, $P_{\theta}$ are
  the real space 
galaxy  power 
spectra of galaxies, galaxy-$\theta$ and $\theta$, respectively;
  $\sigma_v$ is the 1D velocity dispersion; and $F(x)$ is a 
smoothing function, normalized to unity at $x=0$, determined by the
  velocity probability distribution.  This simple formula  has
passed tests in simulations on scales where  $\delta\la 1$
  \cite{Scoccimarro04}. The 
derivation of Eq. (\ref{eqn:RD}) is quite general, so it should be applicable
  even when gravity is modified.

 The distinctive dependence of $P^s_g$ on $u$
  allows for simultaneous determination of $P_g$, $P_{g\theta}$ and
  $P_{\theta}$~\cite{Tegmark02}. 
The  parameters we want to determine are the band powers of $P_{g\theta}(k)$
  \footnote{
 Distance $D$ and $H$ are required to translate the
    observed galaxy angular and redshift separation to ${\bf k}$. In general, errors in $D$ 
    and $H$ measurements cause both horizontal and vertical shifts in the $E_G$
    plot.  Both $D$ and $H$ will be
  measured by methods like type Ia supernovae and baryon acoustic
  oscillations  with $1\%$ accuracy, much smaller than the $k$ bin size
 adopted, so the horoziontal shift is negligible. 
Errors in $D$ show up  in both $P_{g\theta}$ and the $C_{\kappa g}\rightarrow
    P_{\nabla^2(\psi-\phi)g}$ inversion through $l=kD$ and thus  largely cancel
  in evaluating $E_G$.  Errors in $H(z)$ only show up in
    $P_{g\theta}$ measurement and thus cause a net shift in the value
    of $E_G$. For $1\%$ error in $H$, the fractional error in $E_G$ is
    $\leq (n_{\rm eff}+3)1\%\leq 3\%$. Here, $n_{\rm eff}$ is the
    effective power index of the corresponding power spectra. For the
    fiducial $\Lambda$CDM cosmology, it is
    negative in relevant $k$ range.  Thus errors induced by
    uncertainties in
  $D(z)$ and  $H(z)$ measurement will be sub-dominant, except for SKA,
 which requires better control over systematic errors in
    $D$ and $H$ measurement.  For simplicity,
  we neglect this potential error source.   Measuring $P_{g\theta}$ also requires to
    marginalize over $\sigma_v$. However, in the linear
  regime $k\la 0.2h/$Mpc,  $k^2\sigma^2_v/H^2\ll 1$ and
  $F(k^2u^2\sigma^2_v/H^2)\simeq  1$, for typical value 
    $\sigma_v\sim 300$ km/s. Thus the exact value of $\sigma_v$ is not
    required for our analysis. Without loss of generality, we adopt
    $\sigma_v=300$ km/s.  } defined
  such that    
  $P(k)=P_{\alpha}$ if $k_{\alpha}\leq k<k_{\alpha+1}$, where
  $k_1<k_2<\cdots<k_{\alpha}<\cdots$. We denote 
  $P^{(1)}_{\alpha}$ as the band power of $P_{g\theta}$.  
   For a ${\bf k}_i$ in each $k$ bin, we have a measurement of
   $P^s_g$, which we denote as $P_i$.  The unbiased minimum variance
  estimator of $P^{(1)}_{\alpha}$ is $\hat{P}=\sum W_iP_i$, where
  $W_i=\frac{F_i}{2\sigma^2_i}(\lambda_1+\lambda_2 u_i^2+\lambda_3
  u_i^4)$. Here, $F_i\equiv F(ku_i\sigma_v/H)$, $\sigma^2_i$ is the
  variance of $P_i$ and the three Lagrange multipliers $\lambda_{\alpha}$
   ($\alpha=1,2,3$) is determined by
\ba
{\bf \lambda}=(0,\frac{1}{2},0)\cdot {\bf A}^{-1}\ ;\ A_{mn}=\sum_i
   u_i^{2(m+n-2)}\frac{F_i^2}{2\sigma^2_i}\ .
\ea

{\bf Galaxy-galaxy lensing}.---
Weak lensing is sensitive to the convergence $\kappa$, the
projected gravitational potential along the line of sight:
\be
\label{eqn:kappa}
\kappa=\frac{1}{2}\int_0^{\chi_s}\nabla^2(\psi-\phi)W(\chi,\chi_s)d\chi\ .
\ee
Here, $W$ is the lensing kernel. For a flat universe,
$\chi$, $\chi_s$ are 
the comoving angular diameter distance to the lens and source,
respectively. Eq. \ref{eqn:kappa} is a pure 
geometric result and can be applied to any modified
gravity models where photons follow null geodesics. 

A standard method to recover the lens redshift information
is by the  lensing-galaxy cross correlation. For galaxies in  the
redshift range 
$[z_1,z_2]$,  the resulting cross
correlation  power spectrum under the Limber's approximation is 
\ba
C_{\kappa g}(l)&=&\left(4\int_{\chi_{1}}^{\chi_2}
n_g(\chi)d\chi\right)^{-1} \\ 
&\times &\int_{\chi_1}^{\chi_2}
W(\chi,\chi_s)n_g(\chi)P_{\nabla^2( \psi-\phi)g}(\frac{l}{\chi},z)
\chi^{-2} d\chi \nonumber\\
&\simeq& \frac{W(\bar{\chi},\chi_s)}{4l\Delta \chi}\int_{l/\chi_2}^{l/\chi_1}
P_{\nabla^2( \psi-\phi)g}(k,\bar{z})
dk \nonumber \\
&=&\sum_{\alpha}f_{\alpha}(l)P^{(2)}_{\alpha} \nonumber\ .
\ea
Here, $\chi_{1,2}$ are the comoving angular diameter distance to
redshift $z_{1,2}$ and $\bar{\chi}$ is the mean distance.  
The band power $P^{(2)}_{\alpha}$
of $P_{\nabla^2(\psi-\phi)g}$ is defined at the same $k$ range as
$P^{(1)}_{\alpha}$.    In practice, we measure the band power
$C_{\kappa g}(l,\Delta l)$, centered at $l$ with band width $\Delta
l$. The weighting $f_{\alpha}(l,\Delta l)$ is defined 
correspondingly. For each $l$, only a fraction of $\alpha$ having
$f_{\alpha}(l,\Delta l)\neq 0$ contribute.

\begin{figure}
\includegraphics[width=9cm]{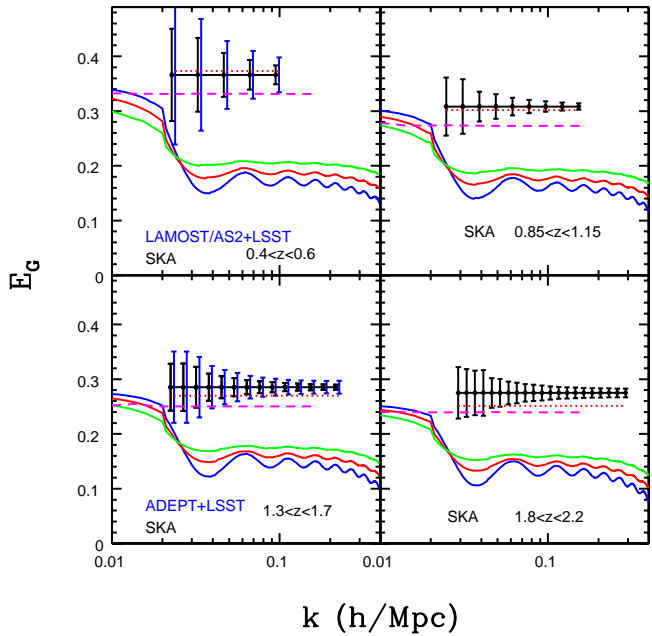}
\caption{$E_G$ as a smoking gun of gravity. Error estimation is
  based on $\Lambda$CDM and error  bars are centered on the $\Lambda$CDM
  prediction (black solid straight line). We only show  those $k$ modes
  well in the linear regime. 
For clarity,
  we shift the error bars of LAMOST/AS2+LSST and ADEPT+LSST slightly
  rightward. 
Irregularities in the error-bars are caused
  by irregularities in the available discrete ${\bf k}$ modes of
  redshift distortion.  Dotted lines are the 
  results of a flat DGP model with $\Omega_0=0.2$. Dashed lines are for
  $f(R)=-\lambda_1 H_0^2\exp(-R/\lambda_2H_0^2)$ with
  $\lambda_2=100$. Differences in expansion histories of these models 
  are of percent level at $z<2$ and are not the main
  cause of differences in $E_G$.  Solid
  lines with wiggles are for TeVeS with $K_B=0.08,0.09,0.1$, where the
  lines with most significant wiggles have $K_B=0.1$. 
 \label{fig:error}} 
 \end{figure}

{\bf A discriminating probe of gravity}.---
With the above measurements, one can
construct an estimator
\be\label{eqn:EGdef}
\hat{E}_G= \frac{C_{\kappa g}(l,\Delta l)}{3H_0^2a^{-1} \sum_{\alpha}
  f_{\alpha}(l,\Delta l)P_{\alpha}^{(1)}}\ ,
\ee
whose expectation value is
\be\label{eqn:EGexp}
\langle \hat{E}_G\rangle=\left[\frac{\nabla^2(\psi-\phi)}{-3H_0^2a^{-1}
  \theta}\right]_{k=\frac{l}{\bar{\chi}},\bar{z}}
=\left[\frac{\nabla^2(\psi-\phi)}{3H_0^2a^{-1} \beta \delta}
\right]_{k=\frac{l}{\bar{\chi}},\bar{z} }\equiv E_G\ .
\ee
The fractional error on
$\hat{E}_G$ is  
\be
\frac{\langle \Delta E_G^2\rangle}{E_G^2}\simeq \frac{\Delta
  C^2}{C_{\kappa g}^2}+\frac{\sum_{\alpha} f^2_{\alpha}\langle (\delta
  P^{(1)}_{\alpha})^2\rangle}{(\sum_{\alpha} f_{\alpha}P^{(1)}_{\alpha})^2}\ ,
\ee
where $\Delta C^2=[C_{\kappa
    g}^2+(C_{\kappa}+C_{\kappa}^N)(C_g+C_g^N)]/(2l\Delta 
  l f_{\rm sky})$. 
Here, $C_{\kappa}$, $C_{\kappa}^N$, $C_g$, $C_g^N$ are the power
spectra of weak lensing convergence, weak lensing shot noise, galaxy
and galaxy shot noise, respectively, and $f_{\rm sky}$ is the fractional sky
coverage. Errors on $E_G$ at any two adjacent bins are correlated, since they
always share some same $k$ modes. 
However,  by requiring
$l_{\alpha}/\chi_1=l_{\alpha+1}/\chi_2$, where
$l_1<l_2<\cdots<l_{\alpha}<\cdots$ and $k_{\alpha}=l_{\alpha}/\chi_2$,   $E_G$
measurement at each $l$ bin  only involves two $k$ bins and thus only errors in
adjacent bins 
are correlated.

\begin{table}
\caption{Summary of target surveys.} 
\begin{tabular}{llllll}
& redshift& deg$^2$&$N_{\rm gal}$&band&operation \\
\hline
\hline
LAMOST\footnote{http://www.lamost.org/en/} &z$<0.8$&10,000&$\sim
10^6$ &optical& 2008 \\
AS2\footnote{Private communication with Daniel Eisenstein}
&z$<0.8$&10,000&$\sim 10^6$ &optical& $\geq 2009$\\
ADEPT\footnote{http://www7.nationalacademies.org/ssb/BE\_Nov\_2006\_bennett.pdf}  
&$1<z<2$ \ \ \ &28,600& $\sim 10^8$ & infrared&$\geq$ 2009\\
SKA\footnote{http://www.skatelescope.org/} & $z\la 5$ & $ 22,000$\
\ \ \ & $\sim10^9$ &radio&2020\\
\hline
LSST\footnote{http://www.lsst.org} & $z\la 3.5$ &10,000&
$\sim 10^9$ &optical&2012
\end{tabular}
\end{table}

We choose ongoing/proposed spectroscopic surveys LAMOST, AS2, ADEPT  and
SKA as targets of
redshift distortion measurements, and  LSST and SKA as targets
of lensing map reconstruction. SKA lensing maps can be  constructed through
cosmic magnification utilizing its unique flux dependence, with S/N
comparable to  that of LSST through cosmic
shear \cite{Zhang05}. Survey specifications are summarized in
TABLE I.  The fiducial cosmology adopted is the  $\Lambda$CDM 
 cosmology, with the WMAP best fit parameters
 $\Omega_0=0.26, \Omega_{\Lambda}=1-\Omega_0, h=0.72,
 \sigma_8=0.77$ and $n_s=1$. The result is shown in figure
\ref{fig:error}. In general, at $k<0.1 h/$Mpc, cosmic variance in
$C_{\kappa g}$  and $P_{g\theta}$  measurements dominates the  $E_G$
error budget, resulting in decreasing error-bars toward larger
$k$. This makes  $f_{\rm  
  sky}$ and the lensing source redshifts the two most relevant survey
parameters for $E_G$ error estimation. Since systematic errors in 
LSST photometric redshifts can be controlled to  better than $1\%$,
errors in $E_G$ measurements of LAMOST/AS2+LSST and ADEPT+LSST
caused by source redshift uncertainties are sub-dominant.

We restrict our discussion to sub-horizon scale perturbations and express 
equations hereafter in the Fourier form. 
Four independent linear equations are required to solve for four perturbation
variables $\delta$, $\theta$, $\psi$ and $\phi$. 
The mass-energy conservation provides two: $\dot{\delta}+H\theta=0$ and
$\dot{H\theta}+2H^2\theta-k^2\psi/a^2=0$.  For at least $\Lambda$CDM,
quintessence-CDM,  DGP and $f(R)$ gravity, the other two takes the general
form 
 \ba
\label{eqn:phipsi}
\phi&=&-\eta(k,a)\psi \ ,\nonumber \\
k^2(\phi-\psi)&=&3H_0^2\Omega_0a^{-1}\delta \tilde{G}_{\rm eff}(k,a) \ .
\ea
Here $\Omega_0$ is the cosmological matter density in unit of the
critical density $\rho_c\equiv 3H_0^2/8\pi G$. Refer to
\cite{Uzan06,Caldwell07,Amendola07} for other ways of parameterizations.
MOND has extra scalar and vector perturbations and does not follow the 
general form of Eq. \ref{eqn:phipsi}~
\cite{Skordis06,Dodelson06}.

(1) {\bf ${\Lambda}$CDM}: $\eta=1$, $\tilde{G}_{\rm
 eff}=1$ and $E_G=\Omega_0/\beta$. 
Dynamical dark energy will have large-scale fluctuations
\cite{DEfluctuation}. Furthermore, it may also have
 non-negligible anisotropic stress and is thus
 able to mimic modifications in gravity \cite{Kunz06}.  But, for
 models with large sound  
 speed and negligible anisotropic stress, such as quintessence,
 these are negligible at sub-horizon scales and 
 Eq. \ref{eqn:phipsi} still holds.

(2) {\bf Flat DGP}:  $\eta=[1-1/3\beta_{\rm
    DGP}]/[1+1/3\beta_{\rm DGP}]$, $\tilde{G}_{\rm eff}=1$ \cite{Koyama06} and
    $E_G=\Omega_0/\beta$, where $\beta_{\rm DGP}=1-2r_cH(1+\dot{H}/3H^2)<0$ and
$r_c=H_0/(1-\Omega_0)$. $\Omega_0$ differs from that of $\Lambda$CDM, in order
    to mimic $H(z)$ of $\Lambda$CDM.

(3) {\bf $f(R)$ gravity}:  in the sub-horizon limit, $\tilde{G}_{\rm
 eff}=(1+f_R)^{-1}$ \cite{Zhang06} and $\eta=1$ \cite{Bean06}, 
 with  $f_R\equiv df/dR|_B$ where $B$ denotes the FRW background. This
 falls naturally out of a conformal transformation of the expression
 for $E_{G}$ in the Einstein  frame into the Jordan frame, noting that
 Einstein frame scalar field  fluctuations are negligible on
 sub-horizon scales \cite{Bean06}. 
We numerically solve the
  full perturbation 
  equations in the Einstein frame since it is computationally simpler
  \cite{Bean06} and then conformally transform to the Jordan frame,
  which we choose as the physical frame, 
  evaluating $\beta$ such that $E_G = \Omega_{0} /(1+f_{R})\beta$. 
In the limit that $f_{R}\rightarrow 0$, e.g. for $f(R)\sim
  \lambda_{1}H_{0}^{2}\exp(-R/\lambda_{2}H_{0}^{2})$ \cite{Zhang06}
  with $\lambda_{1}\ll \lambda_{2}$, the evolution is observationally
  equivalent to $\Lambda$CDM.
For modes that entered the horizon prior to matter-radiation equality,
as we consider here, $\beta$, and therefore $E_{G}$, is scale
invariant for IR modifications to gravity, with  $f_{R}>0$.\footnote{Scales
  larger than the horizon at matter-radiation equality are suppressed
  \cite{Bean06} and, if measurable, would have a scale dependent
  increase in the value of $E_{G}$ in comparison to the small scale
  value.}  The scale independence of $E_G$  holds in $\Lambda$CDM,
 Quintessence-CDM and DGP. 
An observed scale-independent deviation
  in  $E_G$ from $\Lambda$CDM could signify a special
  class of modified gravity, as shown in Fig. \ref{fig:error}.

(4) {\bf TeVeS/MOND}. 
Besides the gravitational metric, TeVeS \cite{MOND} 
contains a scalar and a vector
field. These new fields act as sources for the
gravitational potential $\phi$  in the modified Poisson equation 
and can change the evolution of cosmological
perturbations with respect to standard gravity \cite{Skordis06,Dodelson06}. 
We considered a TeVeS model with $\Omega_b = 0.05$,
$\Omega_\nu = 0.17$, $\Omega_\Lambda = 0.78$ 
and we adopted
a choice of the TeVeS parameters that produces a
significant enhancement of the growth factor. 
  The TeVeS $E_G$
is significantly different from the standard $E_G$
(Fig. \ref{fig:error}).\footnote{To simplify the numerical
  treatment of the TeVeS   perturbations equations while retaining a
  good qualitative description of all the 
significant physical effects at the same time,  we
introduced several approximations. Namely we assumed instantaneous
recombination and employed the tight coupling approximation between
baryons and photons at all scales before decoupling; moreover we evolved 
perturbations in the massive neutrino component in a simplified
way by switching off neutrinos perturbations when they were below the 
free steaming
scale and treating them as a fluid above the free streaming
scale. }
It exhibits scale dependence with accompanying baryon acoustic wiggles. 
Both features are due to the vector field fluctuations, which
{\em{play a significant role}} in structure formation
\cite{Dodelson06}. These fluctuations 
decrease toward small scales and cause the scale dependency of
$E_G$. We also checked that they affect the final shape of the acoustic 
oscillations of the other components significantly. As a result, 
oscillations in $\phi$, $\psi$ and $\delta$ do not cancel out perfectly in 
TeVeS when we take the ratio, thus producing the wiggles in $E_G$.

For the four gravity models investigated, differences in $E_G$ are much
larger than observational statistical uncertainties. Planned surveys
are promising to detect percent level deviation from GR and 
should distinguish these modified gravity models
unambiguously. 

At large scales, gravity is the only force determining the 
acceleration of galaxies and dark matter particles. So we assumed no
galaxy velocity bias. As statistical  errors in $E_G$ 
measurements reach the $1\%$ level (Fig.~1), several systematics,
besides the one discussed in footnote 1,
may become non-negligible. 
 One is the  accuracy of the redshift
  distortion 
formula (Eq.~\ref{eqn:RD}), which may be problematic for those modes with large
$u$,  even at very linear scales \cite{Scoccimarro04}. A remedy is
to  exclude  them when extracting $P_{g\theta}$, at the expense
of statistical  accuracy. As discussed before, accuracy of $E_G$ measurement
is dominated by accuracy of $P_{\nabla^2(\psi-\phi)g}$ measurements
and is thus less  affected. 
A  less severe one is the  nonlinear evolution, which becomes
non-negligible where the matter power spectrum
variance $\Delta^2_m\ga 0.1$.  In general relativity, nonlinear
corrections to  density  
and velocity  differ (Fig. 12, \cite{Bernardeau02}).  A direct
consequence is that $E_G$ develops a dependence on the matter 
power spectrum.  Similar effects in modified gravity models are
expected. This can be corrected by high order perturbation 
  calculations, which should work well where
  $\Delta^2_m\la 0.2$.

We thank R. Caldwell, D. Eisenstein, B. Jain, M.
Kunz, J. Ostriker and J.P. Uzan for useful discussions and the
anonymous referees for useful suggestions. PJZ is supported by the
National Science 
Foundation of China  grant 
10533030 and CAS grant KJCX3-SYW-N2. RB's work is supported 
by the National Science  Foundation grants AST-0607018 and PHY-0555216. SD
is supported by the US 
Department of Energy.

\end{document}